\documentclass[%
 reprint,
 amsmath,amssymb,
 aps,
]{revtex4-1}

\usepackage{graphicx}% Include figure files
\usepackage{dcolumn}% Align table columns on decimal point
\usepackage{bm,braket}% bold math and braket
\usepackage{color} 
\usepackage[dvipsnames]{xcolor}

\begin{document}

\title{A Feshbach engine in the Thomas-Fermi regime}

\author{Tim Keller}
\author{Thom\'as Fogarty}
\author{Jing Li}
\author{Thomas Busch}
\affiliation{Quantum Systems Unit, Okinawa Institute of Science and Technology Graduate University, Onna-son, Okinawa 904-0495, Japan}

\date{\today}% It is always \today, today,

\begin{abstract}
Bose-Einstein condensates can be used to produce work by tuning the strength of the interparticle interactions with the help of Feshbach resonances. In inhomogeneous potentials, these interaction ramps change the volume of the trapped gas allowing one to create a thermodynamic cycle known as the Feshbach engine. However, in order to obtain a large power output, the engine strokes must be performed on a short timescale, which is in contrast with the fact that the efficiency of the engine is reduced by irreversible work if the strokes are done in a non-adiabatic fashion. Here we investigate how such an engine can be run in the Thomas-Fermi regime and present a shortcut to adiabaticity that minimizes the irreversible work and allows for efficient engine operation. 
\end{abstract}

\maketitle

\section{Introduction}

Understanding and exploring concepts in quantum thermodynamics is currently a highly active topic with implications for the future development of quantum technologies \cite{OurManSteve}. Within this area, the creation and operation of quantum engines which implement the Otto cycle and use cold atoms as their working medium, has received special attention, as they can be treated instructively and lend themselves to experimental realisation \cite{Campo2015Bang,Beau2016ScalingUp,li2018efficient,Niedenzu2019quantized,Rossnagel2016IonEngine}. 

Similar to classical thermodynamical engines, quantum engines will achieve maximum efficiency if they are run without creating irreversible work. While this can be achieved by  adiabatic evolution, this mode of operation has the drawback that it requires the external parameter changes to be very slow \cite{born1928beweis}. As reliable, fast and simple control is needed e.g. for the development of new technologies \cite{acin2018quantum}, more recently the use of  shortcuts to adiabaticity (STA) has received increased attention \cite{guery2019shortcuts}. Shortcut protocols provide a way to mimic adiabatic evolution in a finite time, mostly by requiring different parameter ramps or additional levels of control. While a large number of shortcuts have been found for single particle systems, shortcuts for interacting many-particle settings are still rare \cite{Diao2018STAFermi,Fogarty2019twoatoms,Kahan2019CompositeSTA}.  However, first experiments have demonstrated the viability of shortcuts protocols for atomic Bose-Einstein condensates (BECs) in the repulsive mean-field limit \cite{schaff2011shortcut,rohringer2015non} and for fermionic systems in the unitary limit \cite{PDeng2018STAFermiExp}.

Utilizing STA protocols to efficiently drive the dynamics through the control of the external potential parameters, such as trapping frequencies, can therefore increase the performance of finite time quantum heat engines  \cite{Campo2015Bang,Abah_2017}. 
Furthermore, recent works have extended this idea to engines that are driven by changing internal parameters of the working medium, such as changes to the interparticle interaction strength \cite{li2018efficient,Chen2019}. In ultracold atoms these are described by a scattering length, which can be controlled experimentally by varying an external magnetic field about a Feshbach resonance \cite{feshbach1958unified,courteille1998observation,chin2010feshbach}. While applying STA protocols to drive interactions is not a trivial task, Li \textit{et al.} showed that it is possible in the case of a bright solitonic BEC which can be frictionlessly compressed and expanded using designed Feshbach pulses \cite{li2018efficient}. Even though this can lead to an efficient Otto cycle, the operational range of the engine was very limited due to the possibility of BEC collapse in the presence of driven attractive interactions \cite{gerton2000BECcollapse,carr2008multidimensional}. 

It is therefore interesting to extend the idea of the Feshbach engine to BECs in the stable Thomas-Fermi regime of large particle numbers and strong repulsive interactions \cite{dalfovo1999theory}. For this we derive in this work a novel interaction ramp that allows for the frictionless compression and expansion of such a Thomas-Fermi BEC in an almost arbitrarily short time, and show that these ramps can act as STAs in a Feshbach engine. 
The presentation is organized as follows: In Sec.~\ref{sec:sta} we introduce a scaling ansatz to derive an interaction ramp for a harmonically trapped $d$-dimensional BEC in the Thomas-Fermi limit which ensures that the system follows an adiabatic path at all times. We then verify that the ramp is working as intended, up to some minimum time, by numerically simulating the dynamics using the full Gross-Pitaveskii equation and comparing it to a non-optimized reference ramp. In Sec.~\ref{sec:feshbach_engine} we show that the shortcut ramp can indeed be used to increase the power and efficiency of the engine and in Sec.~\ref{sec:modulational_instability} we perform a stability analysis to derive the minimum time in which the interaction ramp can be performed before a modulational instability leads to a condensate collapse. 

\section{Shortcut to adiabaticity}
\label{sec:sta}
In this section we derive an interaction ramp for a BEC in the Thomas-Fermi limit that can act as an STA \cite{chen2010fast,guery2019shortcuts} and evaluate its performance in compressing and expanding a BEC compared to a smooth non-optimized reference ramp. 

\subsection{Interaction ramp}

\begin{figure}
\centering
\includegraphics[width=\columnwidth]{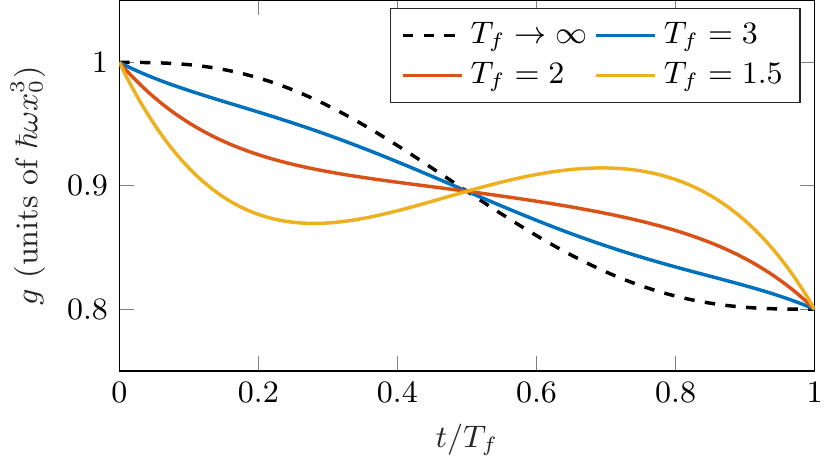}
\caption{Interaction ramps obtained from the shortcut to adiabaticity according to Eq.~\eqref{eq:g_ramp} in three dimensions and for different values of $T_f$. The black dashed line shows the time-rescaled adiabatic reference (TRA). }
\label{fig:ramps}
\end{figure}

For simplicity we start by considering a one-dimensional BEC trapped in an inhomogeneous potential, $V(x)$, and generalise the results to higher dimensions later. The condensate can be described by a single wave function, $\psi(x)$, whose dynamics is governed by the Gross-Pitaevskii equation \cite{pethick_smith_2008}
\begin{equation}
	i\hbar \frac { \partial \psi  } { \partial t } = \left[- \frac { \hbar ^ { 2 } } { 2 m }  \frac{\partial^2}{\partial x^2}  + V ( x )  +  g(t)| \psi | ^ { 2}\right]\psi \, ,
\label{eq:gpe_cc}
\end{equation}
where $m$ is the mass of an individual particle in the condensate, $g(t)$ is the nonlinear interaction strength, which may vary in time, and the wave function is normalized to the number of particles, $\int dx |\psi(x)|^2=N$. Since we are interested in compressing or expanding the width of the wave function without changing its general shape, we choose a scaling ansatz \cite{theocharis2003modulational} of the form
\begin{equation}
	\psi(x,t) = \frac{1}{\sqrt{a(t)}}e^{i\varphi(x,t)}\phi(y(x,t),\tau(t)) \, ,
\end{equation}
where the spatial coordinate is rescaled as $y(x,t) = x/a(t)$ and we have also introduced a rescaled time $\tau(t)$. Inserting this ansatz into the Gross-Pitaevskii equation \eqref{eq:gpe_cc}  and choosing the phase as
\begin{equation}
	\varphi(x,t) = \frac{m}{2\hbar}\frac{\dot{a}(t)}{a(t)}x^2 \, ,
	\label{eq:ChoicePhase}
\end{equation}
to eliminate the term proportional to $\partial \phi/\partial y$, we get
\begin{align}
	i\hbar \frac {\partial \phi } {\partial\tau }\frac{\partial\tau}{\partial t} = &\left[- \frac { \hbar^2} {2m} \frac{1}{a^2}\frac{\partial^2}{\partial y^2}  +V (x) +  \frac{g(t)}{a}| \phi |^2\right]\phi\nonumber\\
		 &+ \left[\frac{i\hbar}{2}\frac{\dot{a}}{a}+\hbar\dot{\varphi}-\frac{i\hbar^2}{2m}\frac{\partial^2\varphi}{\partial x^2} + \frac{\hbar^2}{2m}\left(\frac{\partial\varphi}{\partial x}\right)^2\right]\phi \, .
\end{align}
The choice of phase made in Eq.~\eqref{eq:ChoicePhase} can be interpreted as a gauge transformation $\hat{U}=e^{i\varphi(x,t)}$, which adjusts the momentum of the expanding or shrinking system as
\begin{equation}
	\hat{p} \rightarrow \hat{U}\hat{p}\hat{U}^\dagger = \hat{p} - m\frac{\dot{a}}{a}\hat{x}  \, ,
\end{equation}   
where $\dot{a}x/a$ is the local velocity \cite{castin1996bose,del2013shortcuts}. The same transformation is also commonly found in other scaling problems, e.g.~as the optimal solution in a variational approach \cite{kagan1996evolution,perez1996low,li2016shortcut}. Assuming the external potential is given by a harmonic trap,  $V(x)=\frac{1}{2}m\omega^2x^2$, we obtain
\begin{align}
	i\hbar \frac{\partial \phi} {\partial\tau }\frac{\partial\tau}{\partial t} = \left[- \frac { \hbar ^ { 2 } } { 2 m } \right.& \frac{1}{a^2}\frac{\partial^2}{\partial y^2}\nonumber\\ 
	  + \frac{1}{2}m&\left(\ddot{a}+\omega^2 a\right)ay^2  +\left.  \frac{g(t)}{a}| \phi | ^ { 2}\right]\phi \, .
\label{eq:scaling_gpe}
\end{align}
Choosing the rescaled time $\tau$ and the term $\ddot{a}+\omega^2 a$ in such a way that it leads to a solvable Gross-Pitaevskii equation then allows one to design control pulses for a frictionless evolution of the BEC, e.g.~by varying either the trap frequency $\omega$, the interaction strength $g$ or both \cite{muga2009frictionless,stefanatos2012frictionless,del2013shortcuts,deffner2014classical}. 

Many experiments involving repulsively interacting Bose-Einstein condensates are carried out in the so-called Thomas-Fermi (TF) regime, where the 
potential and the interaction energies are much larger than the kinetic energy, $Ng\gg \sqrt{\hbar^3\omega/m}$ \cite{pethick_smith_2008}. This allows one to neglect the kinetic energy term in the Gross-Pitaevskii equation \eqref{eq:gpe_cc} and obtain an analytical solution of the form
\begin{equation}
	\psi(x,t) = \sqrt{\frac{1}{g}\left(\mu-V(x)\right)}e^{-i\mu t/\hbar} \quad\text{ for }\quad \mu>V(x)
\label{eq:thomas_fermi_wf}
\end{equation}
and $\psi(x,t)\equiv 0$ otherwise.  The chemical potential $\mu$ is determined via the normalization condition. Considering the TF limit 
in the scaling GPE Eq.~\eqref{eq:scaling_gpe} and choosing scaling functions as \cite{ozcakmakli2012shortcuts}
\begin{equation}
\begin{split}
\ddot{a} + \omega^2a &= \omega^2\frac{g(t)}{gi}\frac{1}{a^2} \quad\text{ and }\\  \tau = \int_0^t dt' \frac{g(t')}{g_ia(t')}  &=   \int_0^t dt' \frac{a(t')}{\omega^2}\left(\ddot{a}(t') + \omega^2a(t')\right)\, ,
\end{split}
\label{eq:scaling_functions}
\end{equation}
with some initial interaction strength $g_i$, then gives
\begin{equation}
i\hbar\frac{\partial\phi}{\partial\tau}= \left[\frac{1}{2}m\omega^2y^2  +  g_i | \phi | ^ { 2}\right]\phi \, .
\end{equation}
This again has the aforementioned Thomas-Fermi solution
\begin{equation}
\phi(y,\tau) = e^{-i\mu_i\tau/\hbar}\sqrt{\frac{1}{g_i}\left(\mu_i-\frac{1}{2}m\omega^2y^2\right)}\;,
\end{equation}
with $\mu_i = \left(\frac{9}{32}m\omega^2N^2g_i^2\right)^{1/3}$. Inserting everything back into the scaling ansatz gives us an analytic expression for the evolution of the wave function
\begin{equation}
\begin{split}
\psi(x,t) = &\frac{1}{\sqrt{a(t)}}e^{i\frac{m}{2\hbar}\frac{\dot{a}(t)}{a(t)}x^2}e^{-i\frac{\mu_i}{\hbar}\int_0^tdt'\frac{g(t')}{g_ia(t')}}\times\\
&\sqrt{\frac{1}{g_i}\left(\mu_i-\frac{1}{2}m\omega^2\frac{x^2}{a^2(t)}\right)} \, , 
\end{split}
\label{eq:wf_analytical}
\end{equation}
and choosing a suitable scaling function $a(t)$ then allows us to reverse-engineer an interaction ramp that will take the system along this evolution. For this we rearrange Eq.~\eqref{eq:scaling_functions} for $g(t)$ and get 
\begin{equation}
g(t) = g_i\frac{a^2(t)}{\omega^2}\left(\ddot{a}(t)+\omega^2a(t)\right)\;.
\end{equation}
Choosing appropriate boundary conditions for $a(t)$ of the form
\begin{equation}
\begin{split}
a(0) &= a_i = 1 \, ,\\ 
a(T_f) &= a_f = \left(g_f/g_i\right)^{1/3} \;,\\ 
\dot{a}(0) &=\dot{a}(T_f)=\ddot{a}(0)=\ddot{a}(T_f) = 0 \, ,
\end{split}
\label{eq:tf_a_cond}
\end{equation}
we can drive the system from the Thomas-Fermi ground state at an initial interaction strength $g_i$ to the ground state at a final value $g_f$ in an almost arbitrarily short time $T_f$, while mimicking an adiabatic evolution. It is worth noting that by doing this the system also acquires an additional, but irrelevant phase that depends on $T_f$. The boundary conditions for $a(t)$ can be fulfilled by a fifth-order polynomial $a(t) = a_i + (a_f-a_i)\left[10s^3 -15s^4 + 6s^5\right]$ with $s=t/T_f$, which has the form of a \textit{smoother step}-function \cite{perlin2002improving}. 
This shortcut to adiabaticity is easily generalised to a $d$-dimensional BEC in an isotropic harmonic trap in the Thomas-Fermi limit by driving the interaction strength according to
\begin{equation}
g(t) = g_i \frac{a^{d+1}(t)}{\omega^2}\left(\ddot{a}(t)+\omega^2 a(t)\right)
\label{eq:g_ramp}
\end{equation}
and requiring $a(T_F)=\left(g_f/g_i\right)^{1/(d+2)}$ as well as replacing $g_ia(t')$ with $g_ia^d(t')$ in the time scaling $\tau$ in Eq. \eqref{eq:scaling_functions}. In the following we will use dimensionless units and scale lengths by $x_0=\sqrt{\hbar/m\omega}$, energies by $\hbar\omega$, time in units of $\omega^{-1}$ and interaction strengths by $\hbar\omega x_0^d$.

\subsection{Evaluation}
As a reference for benchmarking the shortcut performance we define a time-rescaled adiabatic (TRA) stroke, which can be obtained by letting $T_f\rightarrow\infty$ or equivalently by setting $\ddot{a}(t)\equiv 0$ in the interaction ramp $g(t)$.  A comparison between the TRA ramp for compressing a three-dimensional Thomas-Fermi BEC from $g_i=1$ to $g_f=0.8$ with STA ramps for varying $T_f$ is shown in Fig.~\ref{fig:ramps}, and one can immediately notice that faster ramps require larger changes in the interaction strength over the duration of the STA.
\begin{figure}
\centering
\includegraphics[width=\columnwidth]{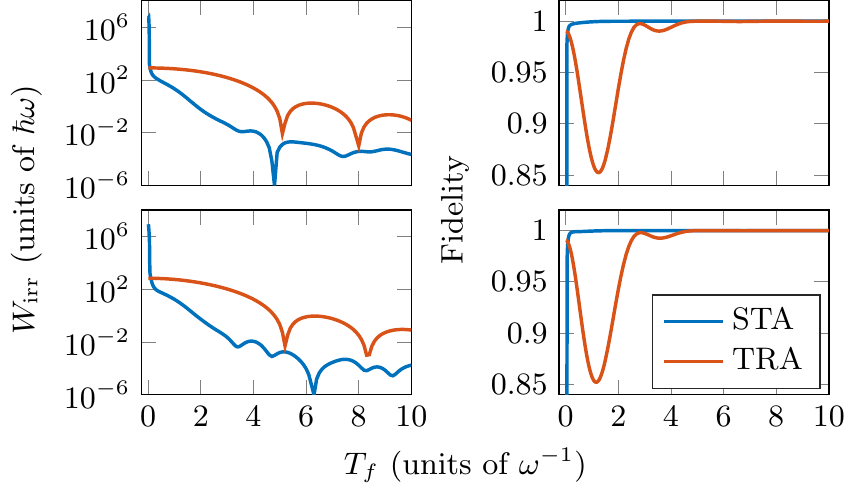}
\caption{\textbf{Upper row:} Irreversible work $W_\mathrm{irr}$ and fidelity $F$ after compressing a 3D BEC consisting of $N=10^4$ atoms from an initial interaction $g_i=1$ to $g_f=0.8$ in a time $T_f$ using the shortcut to adiabaticity (STA) and an adiabatic reference protocol (TRA). \textbf{Lower row:} Identical plots but for the reverse expansion stroke and $N=8000$ atoms.}
\label{fig:strokes}
\end{figure}
We can assess the performance of the shortcut by numerically simulating the Gross-Pitaevskii equation with the calculated interaction ramps and evaluating the irreversible work and fidelity,
\begin{equation}
 W_\mathrm{irr} = E(T_f) - E_f \quad \text{ and } \quad F = \left|\braket{\psi(T_f)|\psi_\mathrm{target}}\right|^2\;,
\end{equation}
at the end of the stroke as a measure of how close the system's state $\ket{\psi(T_f)}$ with energy $E(T_f)$ after the evolution is to the desired target state $\ket{\psi_\mathrm{target}}$ with energy $E_f$. For a three-dimensional BEC consisting of $N=10^4$ atoms  and a compression going from $g_i=1$ to $g_f=0.8$ we show these quantities for the STA and TRA strokes in the upper row of Fig.~\ref{fig:strokes}.  For comparison we also show them for the reverse expansion stroke going from $g_i=0.8$ to $g_f=1$ in the lower row of Fig.~\ref{fig:strokes}, but for $N=8000$ instead. We note we use the actual ground states for the full Gross-Pitaevskii equation as the initial and target states for the evolution, which differ slightly from the Thomas-Fermi wave function in Eq.~\eqref{eq:thomas_fermi_wf}, especially around the condensate edges. Therefore their energy is slightly higher than the Thomas-Fermi value of 
\begin{equation}
E = \frac{5}{7}N\mu \quad\text{ with }\quad \mu = \left(\frac{15Ng}{16\sqrt{2}\pi}\right)^{2/5} \;.
\label{eq:bec_energy}
\end{equation}
Nevertheless, one can see that in each case the STA outperforms the TRA stroke by several orders of magnitude in the irreversible work for almost any stroke duration $T_f$ above some threshold $T_f^\mathrm{min}$. Similarly, above this threshold the shortcut always achieves a perfect fidelity of $F=1$ with the target state while the TRA falls short. The sharp dips in the irreversible work for some stroke times as well as the near-perfect fidelity for the TRA around $T_f\approx\pi$ are accidental and can be attributed to the underlying dynamics in the harmonic trapping potential \cite{haque2013slow}. 

The sudden increase in irreversible work and the accompanying drop of the fidelity to basically zero if the shortcut is performed too fast is due to a modulational instability that exists for the nonlinear Schr\"odinger equation (see for example \cite{sulem2007nonlinear}). It is triggered by the shortcut ramp driving the BEC at attractive interactions for extended periods of time, resulting in a collapse of the condensate. 
This collapse can create trains of bright solitons \cite{strecker2002formation,cornish2006formation,everitt2017observation} 
which lead to a clear deviation from the adiabatic path. In the particular example considered here and in the next section, the threshold below which the modulational instability appears is roughly $T_f^\mathrm{min} \approx 0.05$ and in Section \ref{sec:modulational_instability} we present a more detailed stability analysis to derive a general criterion  for a given dimensionality, chemical potential, change in interaction and stroke time. 

While atom losses due to three-body recombination could hinder performance of this engine cycle, simply limiting the ramp times so that the STA does not need to drive the system at attractive interactions will avoid the resonance point. Similarly, losses close to Feshbach resonances can be largely avoided by decreasing the considered interactions and increasing the number of atoms without leaving the Thomas-Fermi regime, thereby staying far from the resonance point itself. Even with these considerations, our scheme allows for a considerable speed-up of the BEC manipulation while the scattering lengths required experimentally are well within reach for the broad Feshbach resonances found in $^{85}$Rb or $^{7}$Li \cite{cornish2000stable,pollack2009extreme}.

\section{Feshbach engine}
\label{sec:feshbach_engine}
\begin{figure}
\centering
\includegraphics[width=\columnwidth]{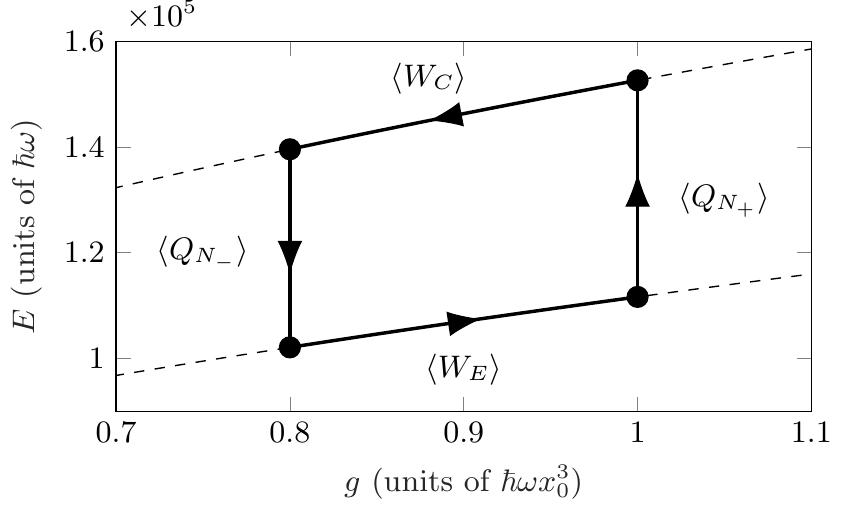}
\caption{The energy of the system in the Thomas-Fermi regime according to Eq.~\eqref{eq:bec_energy}, as a function of the interaction strength $g$ for $N=10^4$ (upper dashed line) and $N=8000$ (lower dashed line). The solid lines indicate the Otto engine cycle consisting of two adiabatic strokes between $g_i=1$ and $g_f=0.8$, performing compression and expansion work, $\langle W_C\rangle$ and $\langle W_E\rangle$ respectively, and two isochoric strokes adding or removing heat $\langle Q_{N_+}\rangle$  or $\langle Q_{N_-}\rangle$ respectively by adding or removing particles to and from the condensate.}
\label{fig:TF_3D_g1_cycle}
\end{figure}
\begin{figure}
\centering
\includegraphics[width=\columnwidth]{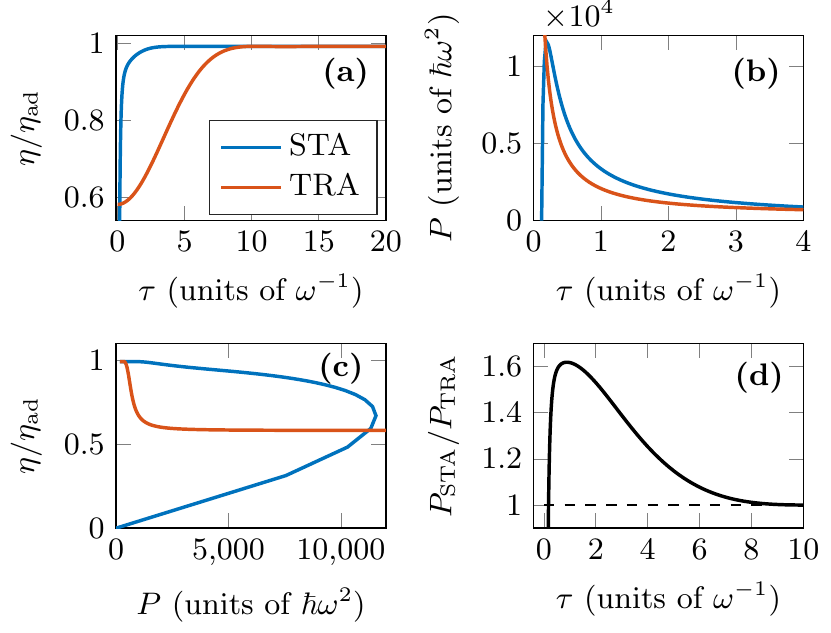}
\caption{(a) Efficiency $\eta$ and (b) power $P$ of the three-dimensional Feshbach engine as a function of the engine cycle duration $\tau = 2T_f$, comparing the performance of the shortcut to adiabaticity (STA) for the adiabatic strokes with an adiabatic reference (TRA). The maximally attainable efficiency for this specific engine cycle is $\eta_{\mathrm{AD}} = 0.0854$. (c) Plot of efficiency against engine power for both the STA and TRA adiabatic strokes and (d) relative increase in engine power by using the STA strokes as a function of cycle duration $\tau$.}
\label{fig:performance}
\end{figure}
A Feshbach engine operates an Otto cycle which consists of two adiabatic interaction ramps that compress and expand a BEC, and which are  connected by two isochoric strokes that add or remove particles \cite{li2018efficient}. The latter can principally be realized by cooling or heating the thermal cloud of atoms surrounding any BEC and thus prompting atoms to condense into or evaporate from the condensate, respectively. 
Within the scope of this work we assume that these isochoric strokes can be performed with perfect fidelity and that their duration can be neglected, so that the engine operation can be evaluated purely as a function of the shortcut performance.
In the following we will evaluate the Otto engine cycle driven in a harmonically trapped, three-dimensional BEC, where the  interaction strength goes between $g_i=1$ and $g_f=0.8$ for the adiabatic strokes and the particle number between $N_i=10^4$ and $N_f=8000$ for the isochoric strokes (see Fig.~\ref{fig:TF_3D_g1_cycle}). We quantify the performance of the engine by calculating its efficiency and power
\begin{equation}
\eta = -\frac{\langle W_C \rangle + \langle W_E\rangle}{\langle Q_{N_+}\rangle} \quad \quad P = -\frac{\langle W_C \rangle + \langle W_E\rangle}{\tau} \, ,
\end{equation}
where the compression and expansion work $\langle W_C\rangle=E(N_i,g_f)-E(N_i,g_i)$ and $ \langle W_E\rangle = E(N_f,g_i)-E(N_f,g_f)$ as well as the heat $\langle Q_{N_+}\rangle = E(N_i,g_i)-E(N_f,g_i)$ are always calculated from the actual values obtained after performing the strokes and not from the adiabatic values indicated in Fig.~\ref{fig:TF_3D_g1_cycle}. After neglecting the dynamics of the isochoric strokes we find that the cycle duration is $\tau\approx 2T_f$ with the work stroke duration $T_f$.

An analytical expression for the maximally attainable adiabatic efficiency of such a Thomas-Fermi Feshbach engine can be obtained by using the Thomas-Fermi wave functions to calculate the energy of the BEC at the engine cycle end points and is given as a function of the compression ratio $g_f/g_i$ by 
\begin{equation}
\eta_\mathrm{AD} =  1 - \left(\frac{g_f}{g_i}\right)^\gamma \, ,
\end{equation}
with $\gamma = 2/3$ (1D BEC), $\gamma = 1/2$ (2D BEC) and $\gamma = 2/5$ (3D BEC). It is worth noting that for the one-dimensional case this is less efficient than the bright soliton Feshbach engine using the same compression ratio, which has an exponent of $\gamma = 2$ \cite{li2018efficient}. 

The efficiency and power of the engine cycle, using the adiabatic strokes shown in Fig.~\ref{fig:strokes}, are plotted in Fig.~\ref{fig:performance}(a) and (b) respectively. One can immediately see that the STA enables the engine to reach its maximum adiabatic efficiency of $\eta_{\mathrm{AD}} = 0.0854$ already for cycle times four to five times shorter than in the TRA case and leads to a considerable increase in engine power for all cycle times considered, as long as the modulational instability is not triggered. Plotting the ratio $P_\mathrm{STA}/P_\mathrm{TRA}$ in Fig.~\ref{fig:performance}(d) shows that this increase can reach up to $60\%$. As expected, the advantage decreases and the STA and TRA perform equally well once the cycle times are increased to more and more adiabatic values. Finally, plotting the engine efficiency versus power in Fig.~\ref{fig:performance}(c) shows that the shortcut enables the engine to run with high efficiency even at high power output, which is an important factor in the operation of any heat engine \cite{lebon2008understanding}. 

\section{Modulational Instability}
\label{sec:modulational_instability}
To better understand the limit of the TF Feshbach engine, we perform in the following a stability analysis to determine the threshold times, $T_f^\mathrm{min}$, below which the shortcuts to adiabaticity derived in Sec.~\ref{sec:sta} fail due to the appearance of a modulational instability. These threshold times act as an intrinsic quantum speed limit \cite{deffner2017quantum} for the manipulation of the Thomas-Fermi BECs considered here. For simplicity, we will again first perform the analysis for the compression of a one-dimensional BEC and show later that the obtained stability criterion can easily be extended to higher dimensions. This is confirmed by comparison with numerical simulations. 

\begin{figure}
\centering
\includegraphics[width=\columnwidth]{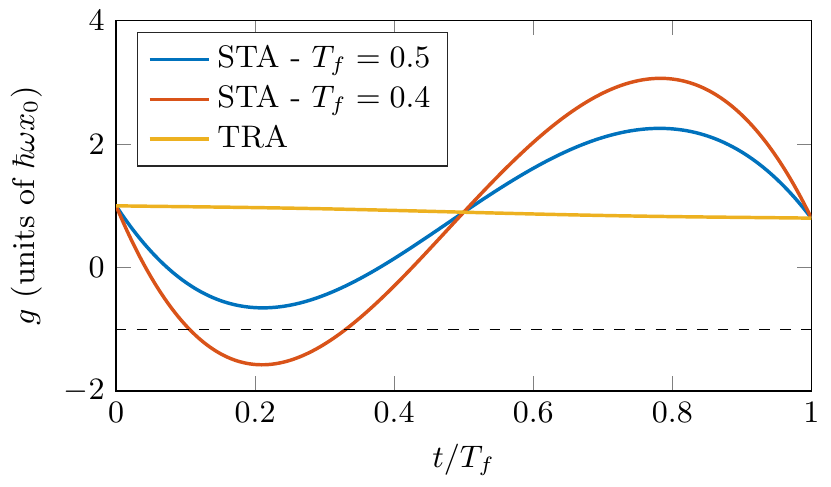}
\caption{Interaction ramp given by the shortcut to adiabaticity for reducing the interaction strength of a one-dimensional BEC from $g_i=1$ to $g_f=0.8$ in time $T_f$ according to Eq. \eqref{eq:g_ramp}. The time-rescaled adiabatic reference (TRA) obtained by setting $\ddot{a}\equiv 0$ is used for comparison. For these parameters, the modulational instability seems to be triggered once the ramp's minimum goes below $-g_i$ (dashed line), which happens roughly around $T_f\approx 0.45$. }
\label{fig:MI_ramp}
\end{figure}

The general reason for an instability to occur is that for short manipulation times $T_f$, the system needs to be driven into the regime of attractive interactions, $g<0$, for a certain amount of time (see Fig.~\ref{fig:MI_ramp}), and entering this regime too deeply will lead the condensate to collapse \cite{carr2004spontaneous}. Using the same parameters of $N=10^4$, $g_i=1$ and  $g_f=0.8$ as in the previous sections, we find that the modulational instability for compressing a one-dimensional BEC via the shortcut is triggered around $T_f^\mathrm{min}\approx 0.45$. 
\begin{figure}
\centering
\includegraphics[width=\columnwidth]{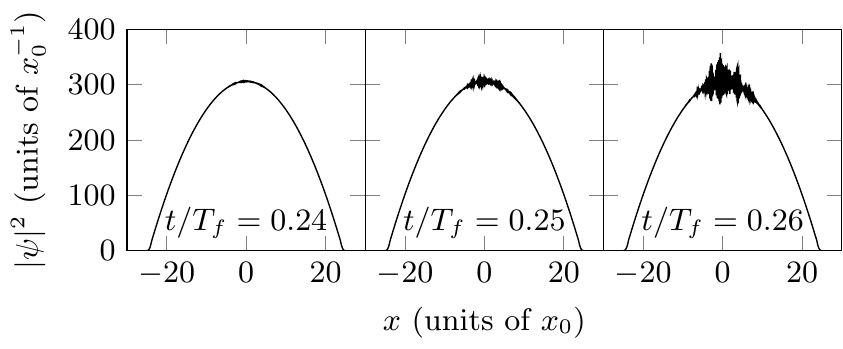}
\caption{Emergence of the modulational instability in the condensate density for compressing a one-dimensional BEC with $N=10^4$ atom in the Thomas-Fermi regime from $g_i=1$ to $g_f=0.8$ by the interaction STA in Eq. \eqref{eq:g_ramp} in a time $T_f=0.4$.}
\label{fig:MI_energy_combined}
\end{figure}

To understand this instability, let us first carefully examine its appearance in the Gross-Pitaevskii dynamics. From the density distributions shown in Fig.~\ref{fig:MI_energy_combined} one can see that the instability manifests itself by the appearance of rapid oscillations that develop in the center of the condensate where the density is maximal. Inserting the analytical solution for the wave function dynamics from Eq. \eqref{eq:wf_analytical} into the GPE gives 
\begin{equation}
i \frac { \partial \psi  } { \partial t } = \left[- \frac { 1 } { 2 }  \frac{\partial^2}{\partial x^2}   -\frac{1}{2}\frac{\ddot{a}(t)}{a(t)}x^2  +  \mu_i\left(a\ddot{a}+a^2\right)\right]\psi \, ,
\label{eq:GPE_stability}
\end{equation}
from which, or by directly setting $x=0$ in Eq.~\eqref{eq:wf_analytical}, one can see that close to the trap center, where the density is approximately homogeneous  and the kinetic and potential energies can be neglected, the wave function evolves according to
\begin{equation}
\psi_\mathrm{hom}(t) = \sqrt{\frac{\mu_i}{a(t)g_i}}e^{-i\int_0^td\tau \mu(\tau)} \;,
\end{equation}
where $\mu(t) = \mu_i(\ddot{a}a + a^2)$. It is worth noting that from Eq.~\eqref{eq:GPE_stability} one can see that the interaction ramp accomplishes the condensate rescaling by creating a time-varying harmonic trapping potential with a time-varying ground state energy.

We assume that the modulational instability can be described by a perturbation to the homogeneous solution of the form of $\psi(x,t) = \psi_\mathrm{hom}(t)\left[1+u(t)\cos(kx)\right]$, with complex amplitude $u(t) = u_\mathrm{re}(t) + iu_\mathrm{im}(t)$ and wave number $k$. Inserting this into the Gross-Pitaevskii equation and keeping only terms up to first order in $u$ then leads to 
\begin{equation}
\frac{\partial}{\partial t}\begin{pmatrix} u_\mathrm{re}\\ u_\mathrm{im} \end{pmatrix} = 
\begin{pmatrix} \frac{\dot{a}}{2a} & \frac{k^2}{2}\\ -\left[\frac{k^2}{2}+2\mu(t)\right] & \frac{\dot{a}}{2a} \end{pmatrix}
\begin{pmatrix} u_\mathrm{re}\\ u_\mathrm{im} \end{pmatrix} + \begin{pmatrix}\frac{\dot{a}}{2a}\frac{1}{\cos(kx)}\\ 0 \end{pmatrix} .
\label{eq:ODE_1}
\end{equation}
Since we are interested in perturbations with a length scale $k\sim\frac{1}{\zeta}$ comparable to the condensate's healing length $\zeta = 1/\sqrt{2\mu}$ \cite{dalfovo1999theory}, all the terms proportional to $\dot{a}/2a$ are negligibly small and we can consider the ODE
\begin{equation}
\ddot{u}_\mathrm{re} +\frac{k^2}{2}\left(\frac{k^2}{2} + 2\mu(t)\right)u_\mathrm{re} = 0 \, ,
\label{eq:ODE_2}
\end{equation}
which is similar to Hill's equation \cite{magnus2013hill}. Note that the same equation can be obtained for the $2D$ and $3D$ case by replacing $\mu_i$ and $a_f$ in $\mu(t)$ with the appropriate values.

Let us note that if $\mu(t)$ was periodic, i.e.~if we had choosen a sinusoidal function to fulfil the boundary conditions \eqref{eq:tf_a_cond} \cite{fogarty2019fast},  Floquet theory would provide exact conditions for the stability of Eq.~\eqref{eq:ODE_2} \cite{magnus2013hill,teschl2012ordinary}. However, a ramp resulting from such an approach generally has maxima and minima with larger magnitude than the corresponding polynomial ramp and therefore triggers the instability even earlier. Furthermore, the instability would usually occur around $t\approx T_f/4$, when the ramp reaches its minimum, which means that the periodicity of the ramp would not come into play. 

\begin{figure}
\centering
\includegraphics[width=\columnwidth]{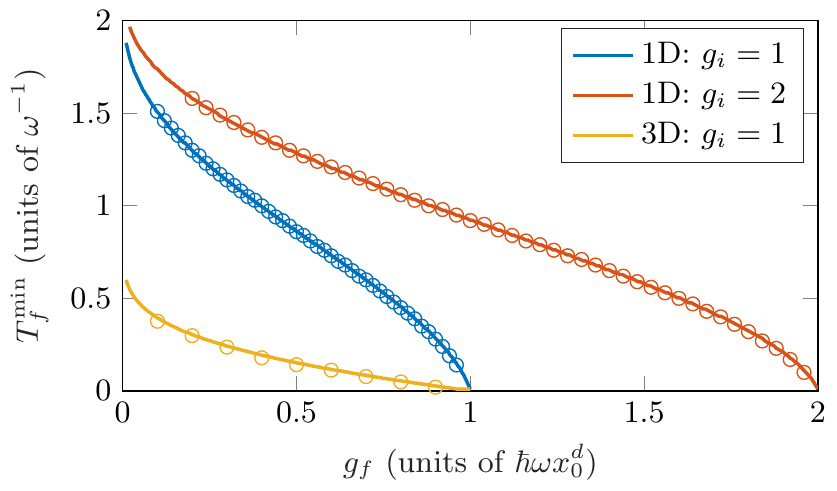}
\caption{Minimum time for the compression of a one-dimensional BEC in the Thomas-Fermi regime consisting of $N=10^4$ atoms by changing its interaction from $g_i=1$ (blue curve) or $g_i=2$ (red curve) to a final value of $g_f$ via the STA in Eq. \eqref{eq:g_ramp}. The curves represent the lines $\Delta=10^{14}$ for the stability criterion \eqref{eq:stability_criterion}. The yellow curve shows $T_f^\mathrm{min}$ for a three-dimensional BEC with $N=10^4$, $g_i=1$ and the same $\Delta = 10^{14}$. The circles show numerically obtained values for $T_f^\mathrm{min}$ from simulating the full Gross-Pitaevskii equation in each case and using the sharp increase in $W_\mathrm{irr}$ as an indicator for the instability.}
\label{fig:TF_stability_analysis}
\end{figure} 
Since we are considering a complex amplitude $u(t)$, it is helpful to do a change of coordinates.  For this we rewrite Eq. \eqref{eq:ODE_1} in polar coordinates using $u(t)=r(t)e^{i\varphi(t)}$ and omitting the terms proportional to $\dot{a}/2a$ to get 
\begin{subequations}
\begin{align}
\dot{r} &= -2\mu(t)\cos(\varphi)\sin(\varphi)r = -\mu(t)\sin(2\varphi)r\;,\\ 
\dot{\varphi} &= -\left[\frac{k^2}{2}+2\mu(t)\cos^2(\varphi)\right]\;.
\end{align}
\end{subequations}
In this form it is easy to see that for $\mu(t)<0$ and $\varphi = \arccos\left(\frac{k}{2\sqrt{|\mu(t)|}}\right)$ we have $\dot{\varphi} = 0 $ and the amplitude of the perturbation can increase exponentially according to
\begin{equation}
\dot{r} = rk\sqrt{|\mu(t)|-\frac{k^2}{4}}\;,
\end{equation}
if $k<2\sqrt{|\mu(t)|}$. The increase is maximal for $k=\sqrt{2|\mu(t)|}$, for which we have
\begin{equation}
\dot{r} = \tilde{\mu}(t) r \quad \text{ or } \quad r(t) = e^{\int_0^td\tau \tilde{\mu}(\tau)}r(0)\;, 
\end{equation}
and where we have defined
\begin{equation}
\tilde{\mu}(t) = \begin{cases} 
      \left|\mu(t)\right| & \text{for } \mu(t) \leq 0 \\
      0 & \text{otherwise}\;.
   \end{cases}
\end{equation}
This allows us to define the total relative increase in the perturbation's amplitude after the interaction ramps as
\begin{equation}
\Delta = \exp\left(\int_0^{T_f} d\tau \tilde{\mu}(\tau)\right) \, .
\label{eq:stability_criterion}
\end{equation}
The previous observation that the modulational instability starts forming at the center of the condensate indicates that it is triggered or seeded by noise \cite{carr2004pulsed,nguyen2017formation}. In  our numerical simulations this is numerical noise, which appears on the level of $10^{-14}$. Since it grows exponentially, a good choice as a criterion for stability is to set $\Delta = 10^{14}$, i.e.~the point at which the small perturbation on the BEC's wavefunction becomes comparable to magnitude of the BEC wavefunction itself.
The resulting curves for a compression stroke are shown in Fig. \ref{fig:TF_stability_analysis}. The minimal compression times $T_f^\mathrm{min}$ from an initial interaction $g_i$ to some final value $g_f<g_i$ agree well with the actual times obtained from numerically simulating the GPE both for one-dimensional and three-dimensional BECs. While we found $\Delta = 10^{14}$ to be the best fit for our numerical data, changing the criterion even by several orders of magnitude only leads to slight deviations in the resulting stability curves. 

\section{Conclusion}
By using a scaling ansatz we have exactly solved the dynamics of an atomic Bose-Einstein condensate subject to a specific interaction ramp in the repulsive Thomas-Fermi limit. This interaction ramp provides a shortcut to adiabaticity for driving the condensate from one interaction strength to another and thereby compressing or expanding it in a short amount of time while avoiding unwanted excitations. We have shown how these shortcuts can increase the efficiency of a Feshbach engine by using them as the adiabatic strokes of its Otto cycle. Using numerical simulations of the full condensate dynamics, we have shown that the speed-up of the condensate manipulation is limited by a modulational instability leading to a condensate collapse. The instability is caused by the need to drive the condensate at increasingly attractive interactions for longer periods of time as the ramp becomes shorter and shorter. Finally, we have performed a stability analysis and determined a criterion that provides an accurate limit $T_f^\mathrm{min}$ for a given initial chemical potential and final interaction strength. 

The Feshbach engine's isochoric strokes offer interesting prospects for further studies. While we have assumed that their dynamics are negligible compared to the adiabatic strokes, this might not readily hold in an experiment. For example, one could model the thermalization process simply via Fourier's law \cite{deffner2018efficiency} or even include the thermal cloud into the model \cite{gardiner2003stochastic} to study its influence on engine performance.
\begin{acknowledgments}
This work has been supported by the Okinawa Institute of Science and Technology Graduate University and
used the computing resources of the Scientific Computing and Data Analysis section. Numerical simulations of the three-dimensional BEC dynamics were performed using the GPUE codebase \cite{schloss2018gpue}.
\end{acknowledgments}

\bibliography{tf_engine_literature}% Produces the bibliography via BibTeX.

\end{document}